\documentclass{aa}  
\usepackage{graphicx}
\usepackage{txfonts}
\usepackage{natbib}
\begin{document} 

   \title{The rotation rates of massive stars}

   \subtitle{How slow are the slow ones?}

   \author{J.O. Sundqvist\inst{1}\and
     S. Sim{\'o}n-D{\'{\i}}az\inst{2,3}\and J. Puls\inst{1}\and
     N. Markova\inst{4} }

   \institute{Universit\"atssternwarte M\"unchen, Scheinerstr. 1,
     81679 M\"unchen, Germany\\ \email{mail@jonsundqvist.com}\and
     Instituto de Astrofisica de Canarias, E-38200 La Laguna,
     Tenerife, Spain\and Departamento de Astrofisica, Universidad de
     La Laguna, E-38205 La Laguna, Tenerife, Spain\and Institute of
     Astronomy with NAO, BAS, P.O. Box 136, 4700 Smolyan, Bulgaria}
                                 
   \date{Received 2013-09-27; accepted 2013-10-14}

 
  \abstract
  {Rotation plays a key role
  in the life cycles of stars with masses above $\sim 8\,M_\odot$.
  Hence, accurate knowledge of the rotation rates of such
  massive stars is critical for understanding their properties and
  for constraining models of their evolution.}
  {This paper investigates the reliability
  of current methods used to derive projected rotation speeds $v \sin
  i$ from line-broadening signatures in the photospheric spectra of
  massive stars, focusing on stars that are not rapidly rotating.}
  {We use slowly rotating magnetic
  O-stars with well-determined rotation periods to test the
  Fourier transform (FT) and goodness-of-fit (GOF) methods typically
  used to infer projected rotation rates of massive stars.}
  {For our two magnetic test stars with
  measured rotation periods longer than one year, i.e., with $v \sin i
  \la 1$\,$\rm km\,s^{-1}$, we derive $v \sin i \approx 40-50$\,$\rm
  km\,s^{-1}$ from both the FT and GOF methods. These severe
  overestimates are most likely caused by an insufficient treatment of the
  competing broadening mechanisms referred to as microturbulence and
  macroturbulence.}
  {These findings warn us not to rely uncritically on results from
  current standard techniques to derive projected rotation speeds of
  massive stars in the presence of significant additional line
  broadening, at least when $v \sin i \la 50$\,$\rm km\,s^{-1}$. This
  may, for example, be crucial for i) determining the statistical
  distribution of observed rotation rates of massive stars, ii)
  interpreting the evolutionary status and spin-down histories of
  rotationally braked B-supergiants, and iii) explaining the
  deficiency of observed O-stars with spectroscopically inferred $v
  \sin i \approx 0$\,$\rm km\,s^{-1}$. Further investigations of
  potential shortcomings of the above techniques are presently under
  way.

}


   \maketitle
%

\section{Introduction}
\label{intro}

Rotation influences all phases of a massive star's life, from the
initial collapse of the rotating molecular cloud
\citep[e.g.,][]{Bodenheimer95}, over the star's main-sequence and
evolved stages \citep[e.g.,][]{Maeder00}, to its violent death as a
core-collapse supernova or hypernova, demises that in some cases may
even produce very energetic bursts of gamma-rays
\citep[e.g.,][]{Yoon06}.

For most stars it is not possible to directly measure the rotation
rate; instead, one typically infers the \textit{projected} stellar
rotation, $v \sin i$ (with inclination angle $i$), from the observed
broadened line-spectrum. A key issue then becomes how to separate this
rotational line broadening from other broadening components present in
the atmosphere. In particular, it is well known that the occurrence of
large so-called macroturbulence seriously complicates deriving
accurate $v \sin i$ rates for massive stars that are not rotating too
rapidly \citep{Conti77, Howarth97, Ryans02, SimonDiaz07, Lefever07,
  SimonDiaz10b, Markova11}.

Several techniques for separating the effects of such ``turbulent''
and rotational broadening in massive stars have been developed and
applied \citep[e.g.,][]{Ryans02, SimonDiaz07}. For example, the
Fourier transform (FT) method uses the first zero in the Fourier power
spectrum to directly extract $v \sin i$, and the goodness-of-fit (GOF)
method simply uses the different shapes of rotational and turbulent
broadening to obtain a best fit of the spectral line. However,
primarily because the underlying physics of additional line broadening
in massive stars is still largely unknown (but see \citealt{Aerts09};
\citealt{Cantiello09}; \citealt{SimonDiaz10b}; \citealt{Shiode13};
\citealt{Sundqvist13b}), it is at present not clear to what degree
these commonly used methods can  actually isolate the rotational
component, and so be used to derive quantitatively accurate values of
$v \sin i$.
 
This Letter tests the FT and GOF methods by using high-quality spectra
of slowly rotating O-stars with detected large-scale magnetic fields,
collected within the Magnetism in Massive Stars project (MiMeS,
\citealt{Wade12}). The rotation period of these magnetic massive stars
can be readily obtained from the observed variation of the
longitudinal field \citep[e.g.,][]{Borra80, Bychkov05} or
photometric/spectral variations caused by their circumstellar
magnetospheres \citep[e.g.,][]{Landstreet78, Howarth07}. In
particular, we focus here on two magnetic O-stars with periods longer
than one year, i.e., with $v \sin i \la 1$\,$\rm km\,s^{-1}$.  For
these stars we derive projected rotation speeds using the FT and GOF
methods, and critically evaluate the reliability of both techniques in
the slow-rotation regime of massive stars.


\section{Observations and method}
\label{observations} 

As our test objects, we use the Of?p-stars \citep{Walborn72}
HD\,191612 and HD\,108, the two known magnetic O-type stars with $v
\sin i \, \la$\,1\,$\rm km\,s^{-1}$ and macroturbulent velocities in
excess of 50 $\rm km\,s^{-1}$ (\citealt{Martins12};
\citealt{Sundqvist13b}; Sim{\'o}n-D{\'{\i}}az \& Herrero 2013).  Of
these two stars, HD\,191612 has the shorter rotation period, $P = 538$
days, which along with its estimated stellar radius $14.5 R_\odot$ and
inclination angle $i \approx 35-50^{\circ}$ yields $v \sin i < 1$ km/s
\citep{Howarth07, Sundqvist12b}. For the even longer period of
HD\,108, $v \sin i < 1$ km/s independent of $i$ \citep[see,
  e.g.,][]{Petit13}. Although a few more magnetic O-stars with well
constrained rotation periods also are known \citep[see compilation
  by][]{Petit13}, they all have either shorter periods or weaker
macroturbulence. Shorter periods imply a non-zero and uncertain true
projected rotation speed (due to uncertain stellar radii and even more
so to an often poorly constrained inclination angle), and lower
macroturbulence means a weaker influence of the primary competing
broadening mechanism that complicates the derivation of accurate $v
\sin i$ values. Since the aim of this paper is to test current methods
using stars with known (and very slow) rotation rates, we restrict the
analysis here to HD\,191612 and HD\,108, deferring to a future paper a
complete study of all OB-stars with detected surface magnetic fields
(see also discussion in Sect.~\ref{discussion}).

High-quality spectra are retrieved from the extensive MiMeS database,
and were obtained using the high-resolution ($R\sim65\,000$)
spectropolarimeter ESPaDOnS at the Canada-France Hawaii Telescope
(HD\,191612) and its twin instrument Narval at the T\'elescope
Bernard-Lyot (HD\,108). As in \citet{Sundqvist13b}, we focus on
analyzing the stars' low states, defined according to when the cyclic
H$\alpha$ emission originating from their ``dynamical magnetospheres''
\citep{Sundqvist12b} is at minimum. This should minimize any
contamination of the photospheric lines by magnetospheric
emission. However, for comparison, in Sect.~\ref{results} we also
comment briefly on differences between this low state and the
corresponding high state, defined by maximum H$\alpha$ emission.

Since Zeeman splitting only contributes $\sim$\,1-2\,$\rm km\,s^{-1}$
per kG in the optical, and the measured dipolar field strengths of
HD\,191612 and HD\,108 are 2.5\,kG and $\sim$\,0.5\,kG \citep{Wade11a,
  Martins10}, magnetic broadening is negligible for both stars
\citep{Sundqvist13b}. This allows us to derive projected rotational
velocities by means of the same techniques as for non-magnetic
stars. For this objective, we use the photospheric O\,III
$\lambda$5591line and apply the {\sc iacob-broad} tools developed by
Sim{\'o}n-D{\'{\i}}az \& Herrero (2013). In summary, the FT method
directly estimates $v \sin i$ from the position of the first zero in
Fourier space, whereas the GOF method convolves synthetic line
profiles for a range of projected rotational and macroturbulent
velocities, creating a standard $\chi^2$-landscape from which a best
combination of the two parameters is determined. For simplicity, in
this Letter we focus exclusively on the above mentioned O\,III line
(as in Markova et al. 2013 and Sim{\'o}n-D{\'{\i}}az \& Herrero 2013),
but we have verified that a corresponding analysis of the C\,IV
$\lambda$5801 line gives very similar $v \sin i$ results \citep[see
  also][]{Howarth07}.

To be consistent with most recent studies of non-magnetic massive
stars (e.g., \citealt{SimonDiaz10b}; \citealt{Najarro11}; Markova et
al. 2013a), we adopt the ``radial-tangential'' (RT) formulation of
Gaussian macroturbulence (rather than the isotropic model used by
\citealt{Sundqvist13b}). This RT model has been the primary choice in
such previous work mainly because the FT and GOF methods then give
consistent $v \sin i$ results (see Fig.~\ref{Fig:GOF}, right
panel). The model assumes that large-scale motions occur only radially
and/or tangentially to the stellar surface (here we assume equal
contributions in both directions), resulting in a more
triangular-shaped profile \citep[see][]{Gray05} as compared to
isotropic macroturbulence. Testing has shown that, for the same
observed line profile, the different shapes of the RT and isotropic
models result in characteristic macroturbulent velocities that differ
on average by $\sim 20\, \rm km\,s^{-1}$ (with the RT model yielding
systematically higher values, Markova et al. 2013b), and that the
choice of macroturbulence also affects the derivation of $v \sin i$
(see Figs.~\ref{Fig:GOF} and \ref{Fig:isotropic}, and discussion in
Sect.~\ref{discussion}).


\section{Results}
\label{results} 

\begin{figure}
  \centering
  {\includegraphics[angle=90,width=6.0cm]{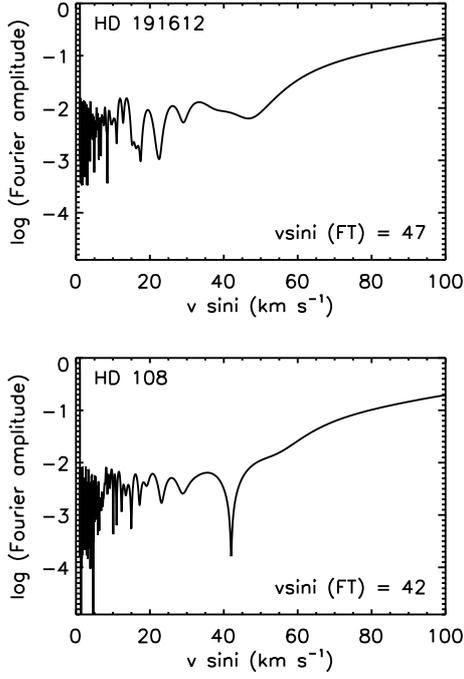}}
  \caption{Fourier power spectra of the O\,III $\lambda$5591 line in
    HD\,191612 and HD\,108. In standard analysis, the position of the
    first minimum is used to infer the projected rotation speed of the
    star.}
  \label{Fig:FT}
\end{figure}

\begin{figure}
    \resizebox{\hsize}{!}
    {\includegraphics[angle=90]{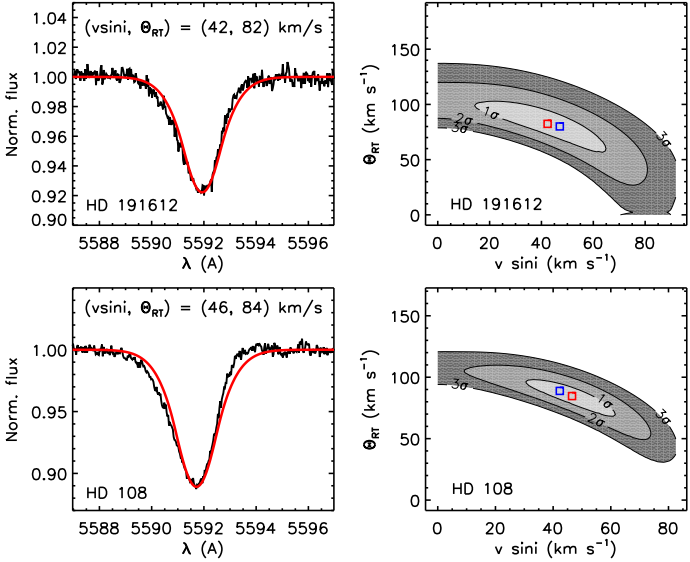}}
    \caption{\textbf{Left:} Best fits of GOF analysis, with derived
      projected rotation speeds and characteristic macroturbulent
      velocities $\theta_{\rm RT}$ as indicated in the
      figure. \textbf{Right:} Corresponding contour maps, with
      1$\sigma$, 2$\sigma$, and 3$\sigma$ confidence intervals
      indicated. The red squares indicate the best GOF and the blue
      squares the $v \sin i$ derived from the Fourier analysis in
      Fig.~\ref{Fig:FT}}.
      \label{Fig:GOF}
\end{figure}

\begin{figure}
    \resizebox{\hsize}{!}
              {\includegraphics[angle=90]{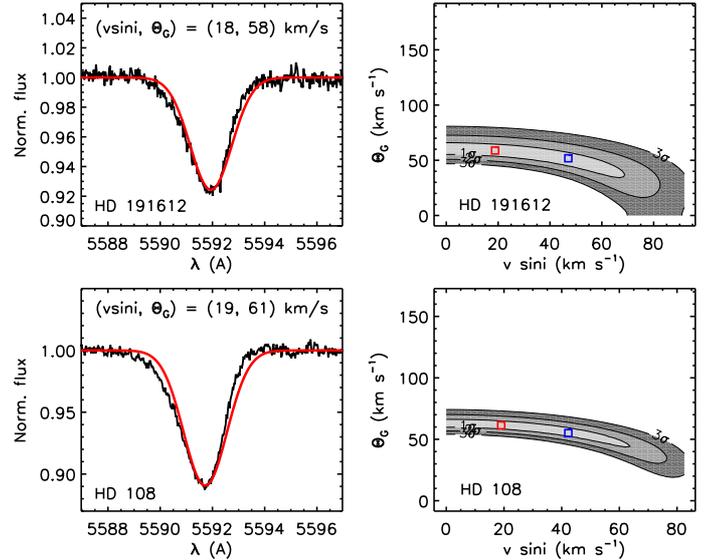}}
  \caption{As Fig.~\ref{Fig:GOF}, but now assuming isotropic
    macroturbulence, with characteristic velocity $\theta_{\rm G}$,
    instead of radial-tangential (see Sect.~\ref{observations}). While
    a ``best fit'' is provided in the left panel, the solutions are in
    principle degenerate in the range $v \sin i \approx 0-60$ $\rm
    km\,s^{-1}$, as illustrated by the contour maps in the right
    panel.}
  \label{Fig:isotropic}
\end{figure}

Figs.~\ref{Fig:FT} and \ref{Fig:GOF} illustrate the derivation of the
projected rotation speeds for HD\,191612 and HD\,108 using the FT and
GOF techniques, respectively. The figures reveal how both methods
yield $v \sin i \approx 40-50$\,$\rm km\,s^{-1}$, a severe
overestimation compared to the $v \sin i \la 1$\,$\rm km\,s^{-1}$
inferred from the variations of the longitudinal magnetic field (see
Sects.~\ref{intro} and \ref{observations})\footnote{We note that while
  the analysis in this paper is based solely on spectra collected by
  the MiMeS collaboration, analysis of complementary {\sc iacob}
  spectra for HD\,191612 and HD\,108 (Sim{\'o}n-D{\'{\i}}az \& Herrero
  2013) gives the same overestimated $v \sin i $ results as found
  here.}.

The GOF contour-maps in Fig.~\ref{Fig:GOF} actually display quite wide
ranges of allowed values for $v \sin i$ and macroturbulent velocities,
where we note in particular that for HD\,191612 $v \sin i \approx 0
\ \rm km\,s^{-1}$ cannot be ruled out with a 2$\sigma$ confidence,
even though the best fit indicates a much higher value. However, from
the first minima in the Fourier transforms displayed in
Fig.~\ref{Fig:FT} we clearly measure projected rotation speeds that
deviate significantly from zero.

\begin{figure}
  \centering
  \includegraphics[width=5.7cm]{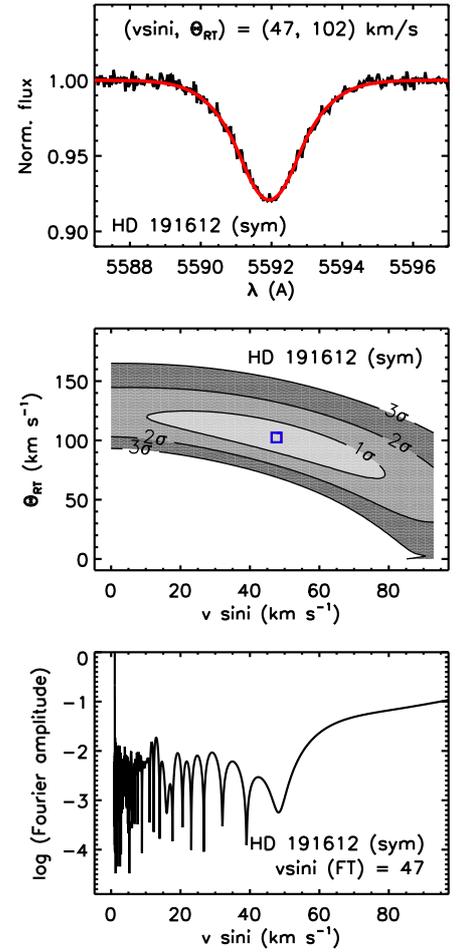}
  \caption{The same as Figs.~\ref{Fig:FT} and \ref{Fig:GOF}, but now
    with the observed O\,III line artificially symmetrized about
    line-center. We note the overlap of the blue and red squares in
    the middle panel, indicating prefect agreement between $v \sin i$
    derived from the FT and GOF methods.}
  \label{Fig:HD191612_sym}
\end{figure}

The line profiles displayed in Fig.~\ref{Fig:GOF} are asymmetric,
which can influence the reliability of $v \sin i$ values, particularly
when inferred from the FT method (e.g., \citealt{Aerts09}). To test
whether this asymmetry might be responsible for the overestimated
projected rotation speeds, we artificially symmetrized the blue wings
of the O\,III lines about their line-centers, and then re-analyzed the
resulting symmetric profiles. However as illustrated for HD\,191612 by
Fig.~\ref{Fig:HD191612_sym}, these symmetric profiles also give a
clear high-velocity minimum in the Fourier power spectra, as well as a
GOF contour-map similar to the one displayed in Fig.~\ref{Fig:GOF},
resulting again in a best estimate $v \sin i \approx 40-50$\,$\rm
km\,s^{-1}$. An equivalent analysis of HD\,108, which is not shown in
the figure, gives similar results, and symmetrizing the red wings of
the profiles also yields highly overestimated projected rotation
speeds, now on the order of $v \sin i \approx 30-40$ km/s (due to the
somewhat narrower redward wings in the original asymmetric profiles).
To further test the possible impact of such line-profile asymmetry and
variability, we also analyzed the O\,III line during HD\,191612's high
state (see Sect.~\ref{observations}), but that fitting yielded
projected rotation speeds on the order of 40 $\rm km\,s^{-1}$ as well.
 
\section{Discussion and conclusions}
\label{discussion}

The central result of this paper is the $v \sin i \approx 40-50$\,$\rm
km\,s^{-1}$ spectroscopically derived for two (magnetic) O-stars with
measured rotation periods implying $v \sin i \la 1$\,$\rm
km\,s^{-1}$. Provided these results can be extrapolated to O-stars in
general (i.e., that there is no additional process that seriously
contaminates the analysis here of the magnetic stars' photospheric
line profiles), the largely overestimated $v \sin i$ values suggest
that, in the presence of strong line broadening in addition to
rotation, neither the FT nor the present implementation of the GOF
method is able to derive reliable projected rotation speeds when $v
\sin i \la 50$ $\rm km\,s^{-1}$. Ongoing (e.g., Sim{\'o}n-D{\'{\i}}az
\& Herrero 2013) and future studies on larger samples will tell
whether the stars examined here indeed are the forerunners of a more
general problem.

Such general shortcomings of the present-day techniques for inferring
low projected rotation speeds may then have consequences for
determining the statistical distribution of observed rotation rates of
massive-star populations in the Galaxy and beyond (e.g.,
\citealt{Hunter08}; \citealt{Dufton13}; \citealt{Ramirez13};
Sim{\'o}n-D{\'{\i}}az \& Herrero 2013). It may also help explain the
apparent deficiency of observed O- and early B-stars with
spectroscopically inferred $v \sin i \approx 0$\,$\rm km\,s^{-1}$, a
long-standing problem (e.g., \citealt{Penny96}; \citealt{Howarth97})
which only for the cases of late O- and B-dwarfs seems to have been
alleviated by the consideration of additional broadening due to
macroturbulence (\citealt{SimonDiaz07}; Markova et al. 2013a,b;
Sim{\'o}n-D{\'{\i}}az \& Herrero 2013). Furthermore, the results here
may be important for interpreting the evolutionary status of massive
B-supergiants, which are slow rotators, but which always seem to
display rotation rates a bit higher than zero, on the order of $v \sin
i \approx 30-40$\,$\rm km\,s^{-1}$ (\citealt{Howarth97}; Markova et
al. 2013a,b). If the rates of these B-supergiants have also been
overestimated, it would (at least for masses $M \ga 35 M_\odot$,
Markova et al. 2013a,b) bring observations in quantitative better
agreement with current (single star) evolution models that predict
these objects at essentially zero rotation speeds
(\citealt{Brott11})\footnote{Resulting from strong wind-induced
  rotational braking when crossing to the cool side of the theoretical
  ``bi-stability jump'' of increased mass loss \citep{Vink10}.}.

The overestimations of projected rotation speeds found for the
magnetic stars here are most likely a result of insufficient
treatments and physics-knowledge of the competing broadening
mechanisms microturbulence and macroturbulence. For example, already
\citet{Gray73} pointed out that classical atmospheric microturbulence
can give Fourier-space minima in addition to those resulting from
rotation; indeed, for relatively strong lines like those examined in
this paper, tests using synthetic profiles show that microturbulent
velocities of $\sim$\,20\,$\rm km\,s^{-1}$ produce Fourier-space
minima at frequencies corresponding to $\sim$\,30-40\,$\rm km\,s^{-1}$
(Sim{\'o}n-D{\'{\i}}az \& Herrero 2013)\footnote{Microturbulent
  velocities of $\sim$\,20\,$\rm km\,s^{-1}$ cannot be ruled out for
  O-type stars, particularly not for O-supergiants
  \citep[e.g.,][]{Massey13}, but seem a bit high for the case of
  B-supergiants \citep[e.g.,][]{Markova08}.}. Moreover, since (at
least for lines with saturated cores) the effect of large
microturbulence on the line shape can be mimicked by a suitable
combination of rotation and macroturbulence (Sim{\'o}n-D{\'{\i}}az \&
Herrero 2013), in these cases it is difficult to obtain accurate $v
\sin i$ rates in the slow-rotation regime also when using the GOF
method. Furthermore, the assumed shape of macroturbulence also affects
the projected rotation speed inferred from a GOF solution
(Sim{\'o}n-D{\'{\i}}az \& Herrero 2013). Indeed, repeating the GOF
analysis in Sect.~\ref{results} but now assuming isotropic, instead of
RT, macroturbulence results in degenerate solutions in the range $v
\sin i \approx 0-60 \, \rm km\,s^{-1}$, as illustrated by the contour
maps in Fig.~\ref{Fig:isotropic}. This is consistent with
\citet{Martins12} and \citet{Sundqvist13b}, who obtained reasonable
line profile fits for HD\,191612 and HD\,108 when assuming zero
projected rotation speeds and this type of isotropic
macroturbulence. In summary, to make further progress it seems clear
that a more realistic treatment of photospheric microturbulence and
macroturbulence in massive stars is urgently needed.

Finally, this paper has focused on the extent to which reliable $v
\sin i$ values of \textit{slowly} rotating massive stars can be
derived by current spectroscopic standard techniques, in the presence
of a significant additional line-broadening component. However, while
for \textit{fast} rotators the importance of such additional
broadening decreases, there are other issues related to the derivation
of accurate $v \sin i$ rates in that parameter range (like continuum
normalization of shallow lines; see Sim{\'o}n-D{\'{\i}}az \& Herrero
2013). We thus plan to expand the study here to test the FT and GOF
methods also for such faster rotators, using the significant number of
relatively fast rotating magnetic B-stars with measured rotation
periods \citep[see][]{Petit13} present in the MiMeS database.

\begin{acknowledgements}
 JOS gratefully acknowledges support from DFG grant Pu117/8-1.  The
 MiMeS project is thanked for providing the observed spectra of
 HD\,108 and HD\,191612. SS-D acknowledges financial support from the
 Spanish Ministry of Economy and Competitiveness (MINECO) under the
 grants AYA2010-21697-C05-04, Consolider-Ingenio 2010 CSD2006-00070,
 and Severo Ochoa SEV-2011-0187, and by the Canary Islands Government
 under grant PID2010119.
\end{acknowledgements}

\bibliographystyle{aa}
\bibliography{sundqvist_rot}

\end{document}